# Bremsstrahlung from Charged Bose-Einstein Condensates

**Mark P. Davidson**
Spectel Research Corporation
807 Rorke Way
Palo Alto, CA 94303
mdavid@spectelresearch.com
www.spectelresearch.com

May 1, 2004

**ABSTRACT**

Low energy bremsstrahlung formulae are derived for a particle beam of charged bosons forming a Bose-Einsten condensate. The expression for energy radiated consists of two terms in this case. One of them, the larger one in the limit of large numbers of Bosons in the state, is proportional to the number of bosons squared and has the same form as one obtains in a hydrodynamic model of quantum wave mechanics. This term is sensitive to the size and shape of the wave packet especially when the force field causing acceleration is localized in extent. The second term, identical to the single particle scattering formula, is less sensitive to the wave packet size and shape and is proportional to the number of bosons in the condensate. The conclusion is that for a Bose-Einstein condensate the radiated bremsstrahlung is a sensitive function of the wave packet shape which is quite different than for a beam of incoherent particles which do not show very much dependence on the wave packet. Only lowest order radiation is calculated. It is also found that to lowest order there does not exist any situation in which the radiation loss from a coherent state of bosons vanishes completely, so that there is no analog of superconductivity in a particle beam of this type at least within the net of assumptions made in this paper. The amount of radiation can be greatly suppressed however, as it is a sensitive function of the form of the wave function.

## 1. INTRODUCTION

Bremsstrahlung and synchrotron radiation have been studied extensively in the physics literature[1-12]. This paper addresses the modifications that must be made to the usual formulae to handle charged Bose-Einstein condensates undergoing acceleration. As a first paper on this subject, we consider only non-relativistic particles and spin is ignored.

Bose-Einstein condensates are a topic of current interest in physics [13-18]. Although neutral Bose-Einstein condensates have been the primary focus experimentally, there is theoretical interest in charged condensates as well [19-31]. It is undoubtedly just a matter of time before charged condensates are formed in a laboratory. The general problem of scattering is modified when considering matter waves with extended coherence [32], and the results we derive are consistent with this.

Why study Bremsstrahlung? First of all, once charged condensates are formed it will be natural to try and produce a condensate beam in an accelerator, and the phenomenon of bremsstrahlung will then become an important issue. Bremsstrahlung will lead to decoherence of the beam, and consequently it should be reduced to as small a value as possible in order to preserve the lifetime of the condensate. The radiation problem is interesting because there may be situations where a Bose-Einstein condensate in an accelerator beam might act as a superconductor, or at least have very reduced radiative loss. This would constitute a new form of superconductivity in a gaseous condensate. There would be practical applications in the form of particle beam storage, information storage, and possibly even energy storage if such a superconducting mode for a particle accelerator could be found. Our results do not show superconductivity, but they reveal some interesting behaviour already at lowest order perturbation in non-relativistic QED. Higher order corrections become more important as the mean number of particles in the condensate grows. One important phenomenon which could be important is reabsorption of emitted photons.

## 2. EMISSION FORMULAE

We wish to calculate the bremsstrahlung of charged bosons which are in a Bose-Einstein condensate. We must pay more careful attention to the wave function of the bosons than is usually done in bremsstrahlung calculations, because the condensate exhibits much stronger dependence on the wavefunction's form than does a single particle in bremsstrahlung calculations.

It is common to describe a condensate of neutral bosons by a coherent state in which the common wave function for the particles is given by the Gross-Pitaevskii equation [33, 34]

$$\left[\frac{1}{2m}\left(\mathbf{p}-\frac{e}{c}\mathbf{A}_E\right)^2 + c|\Psi|^2 + V\right]\Psi = i\hbar\frac{\partial \Psi}{\partial t}, \quad \mathbf{p} = -i\hbar\nabla \qquad (1)$$

Where V is an external potential function and $\mathbf{A}_E$ an external vector potential. For high densities this Hamiltonian must be modified to include the long-range interparticle Coloumb interaction, but we will ignore this interaction here. It is usually assumed that the charged particle condensate occurs in a neutralizing background of opposite charge so that overall charge neutrality can be maintained, but in a particle beam if the density of the beam is low enough then the interparticle interaction should be ignorable anyway. The inclusion of a neutralizing background has no effect on the calculations here in any event and so will be ignored. We use periodic boundary conditions applied to a 3 dimensional cube with side L. We can write the mean c number field or wave function as a time dependent plane wave expansion

$$\Psi(x,t) = \frac{1}{L^{3/2}}\sum_{\mathbf{k}} \varphi_{\mathbf{k}}(t)\exp(i\mathbf{k}\cdot\mathbf{x}) \qquad (2)$$

The normalization condition is

$$\sum_{\mathbf{k}} |\varphi_{\mathbf{k}}(t)|^2 = 1 \tag{3}$$

In terms of the boson creation and annhilation operators

$$[\hat{b}_{\mathbf{k}}, \hat{b}_{\mathbf{k'}}^{\dagger}] = \delta_{\mathbf{kk'}}; [\hat{b}_{\mathbf{k}}, \hat{b}_{\mathbf{k'}}] = [\hat{b}_{\mathbf{k}}^{\dagger}, \hat{b}_{\mathbf{k'}}^{\dagger}] = 0 \tag{4}$$

The n particle Fock state vector with this mean field can be written in the Schrödinger representation as

$$|\psi, n, t\rangle = \frac{1}{\sqrt{n!}} \left[ \sum_{\mathbf{k}} \varphi_{\mathbf{k}}(t) (\hat{b}_{\mathbf{k}}^{\dagger}) \right]^n |0\rangle \tag{5}$$

And the coherent state with mean particle number N as

$$|\psi_c, z, t\rangle = \exp(-|z|^2/2) \exp(z \sum_{\mathbf{k}} \varphi_{\mathbf{k}}(t) \hat{b}_{\mathbf{k}}^{\dagger}) |0\rangle = \prod_{\mathbf{k}} \exp(z \varphi_{\mathbf{k}}(t) \hat{b}_{\mathbf{k}}^{\dagger}) |0\rangle, \quad N = |z| \tag{6}$$

The coherent state so defined is an eigenvector of all the $\hat{b}_{\mathbf{k}}$

$$\hat{b}_{\mathbf{k}} |\psi_c, z, t\rangle = z \varphi_{\mathbf{k}}(t) |\psi_c, z, t\rangle \tag{7}$$

The norm of the coherent state is

$$\langle \psi_c, z, t | \psi_c, z, t \rangle = \exp(-|z|^2) \exp(\sum_{\mathbf{k}} |z \varphi_{\mathbf{k}}(t)|^2) = 1 \tag{8}$$

The field operator is

$$\hat{\Phi}(x) = \frac{1}{L^{3/2}} \sum_{\mathbf{k}} \exp(i\mathbf{k} \cdot x) \, \hat{b}_{\mathbf{k}} \tag{9}$$

And it satisfies

$$\left[ \hat{\Phi}(x), \hat{\Phi}(x')^{\dagger} \right] = \delta^3(x - x') \tag{10}$$

The coherent state is an eigenvector for the field operator

$$\hat{\Phi}(x)|\psi_c,z,t\rangle = z\psi(x,t)|\psi_c,z,t\rangle \tag{11}$$

The electromagnetic field operators can be expanded as follows. Working in the Coulomb gauge ($\nabla \cdot \mathbf{A} = 0$)

$$\hat{\mathbf{A}}(\mathbf{x}) = \sum_{\mathbf{k}\lambda} \varepsilon_{\mathbf{k}\lambda} \frac{1}{L^{3/2}} \left[ \exp(i\mathbf{k}\cdot\mathbf{x})\hat{a}_{\mathbf{k}\lambda} + \exp(-i\mathbf{k}\cdot\mathbf{x})\hat{a}_{\mathbf{k}\lambda}^\dagger \right] \tag{12}$$

where $\hat{a}_{\mathbf{k}\lambda}$ is the annihilation operator for the field mode and $\hat{a}_{\mathbf{k}\lambda}^\dagger$ the creation operator satisfying.

$$\left[\hat{a}_{\mathbf{k}\lambda}, \hat{a}_{\mathbf{k}'\lambda'}^\dagger\right] = \frac{2\pi\hbar c}{k} \delta_{kk'}\delta_{\lambda\lambda'} \tag{13}$$

And where $\varepsilon$ denotes the polarization state.

The Hamiltonian for radiation interacting with the boson field can be taken to be

$$\hat{H} = \frac{1}{2m} \int d^3x\, \hat{\Phi}^\dagger(\mathbf{x}) \times \\ \left[ (\mathbf{p} - \frac{q}{c}\mathbf{A}_E(x) - \frac{q}{c}\hat{\mathbf{A}}(x))^2 + V(x) + c\Psi^*(x)\Psi(x) \right] \hat{\Phi}(\mathbf{x}) + \hat{H}_{EM} \tag{14}$$

Where $H_{EM}$ is the purely electromagnetic part of the Hamiltonian and where V(x) includes any external electromagnetic potential. The relevant interaction term of the Hamiltonian is the part linear in $\hat{\mathbf{A}}(x)$ which after an integration by parts can be written

$$\hat{H}_{rad} = \frac{-q}{mc} \int d^3x\, \hat{\Phi}^\dagger(\mathbf{x})\hat{\mathbf{A}}(x)\cdot(\mathbf{p} - \frac{q}{c}\mathbf{A}_E(x))\hat{\Phi}(\mathbf{x}) \tag{15}$$

In the Born approximation this interaction leads to single photon emission and absorption.

We wish to compute the transition amplitude for emission of a photon of wave vector **k** and polarization ε.

The Hamiltonian is of the form

$$\hat{H} = \hat{H}_0 + \hat{H}_{Rad} \tag{16}$$

It will not depend explicitly on time if the external potentials are static which we assume. We assume further that the coherent states satisfy the Schrödinger equation for the Hamiltonian $H_0$ at least approximately

$$\hat{H}_0 |\Psi_c, z, t\rangle = i\hbar \frac{\partial}{\partial t} |\Psi_c, z, t\rangle \tag{17}$$

We wish to calculate the perturbed state which satisfies

$$i\hbar \frac{\partial}{\partial t} |\chi_c, z, t\rangle = \left( \hat{H}_0 + \hat{H}_{rad} \right) |\chi_c, z, t\rangle \tag{18}$$

In the Dirac or interaction representation this becomes

$$i\hbar \frac{\partial}{\partial t} |\chi_{cD}, z, t\rangle = e^{i\hat{H}_0 t} \hat{H}_{rad} e^{-\hat{H}_0 t} |\chi_{cD}, z, t\rangle \tag{19}$$

And we may integrate this to obtain

$$i\hbar |\chi_{cD}, z, t\rangle = i\hbar |\chi_{cD}, z, 0\rangle + \int_0^t dt' e^{i\hat{H}_0 t'} \hat{H}_{rad} e^{-\hat{H}_0 t'} |\chi_{cD}, z, t'\rangle \tag{20}$$

The lowest order perturbation approximation is obtained by iteration one time

$$|\chi_{cD}, z, t\rangle = |\chi_{cD}, z, 0\rangle - \frac{i}{\hbar} \int_0^t dt' e^{i\hat{H}_0 t'} \hat{H}_{rad} e^{-\hat{H}_0 t'} |\chi_{cD}, z, 0\rangle \tag{21}$$

We can assume that t=0 occurs before the interaction gets turned on which would be the case if the external force is localized and the wave packet has not yet arrived at this point yet. Therefore the perturbed and unperturbed states are the same at t=0

$$|\chi_{cD}, z, 0\rangle = |\Psi_c, z, 0\rangle \tag{22}$$

And so we have

$$|\chi_{cD}, z, t\rangle = |\Psi_c, z, 0\rangle - \frac{i}{\hbar} \int_0^t dt' e^{i\hat{H}_0 t'} \hat{H}_{rad} e^{-\hat{H}_0 t'} |\Psi_c, z, 0\rangle \tag{23}$$

Now the Hilbert space is a direct product of the boson Fock space with the photon Fock space. The unperturbed state has no photons, and we are interested in final states which have one photon. The probability amplitude for producing a photon (with wavevector **k** and polarization ε) and some arbitrary boson state by time t is

$$\langle \Psi_{out}, \{\mathbf{k}, \varepsilon\} | \chi_c, z, t \rangle = -\langle \Psi_{out}, \{\mathbf{k}, \varepsilon\} | \frac{i}{\hbar} \int_0^t dt' e^{i\hat{H}_0 t'} \hat{H}_{rad} e^{-\hat{H}_0 t'} |\Psi_c, z, 0\rangle \tag{24}$$

The total probability of producing such a photon is obtained by summing over all the final boson states

$$P(k, \varepsilon) = \sum_{\Psi_{out}} \left| \langle \Psi_{out}, \{\mathbf{k}, \varepsilon\} | \frac{1}{\hbar} \int_0^t dt' e^{i\hat{H}_0 t'} \hat{H}_{rad} e^{-\hat{H}_0 t'} |\Psi_c, z, 0\rangle \right|^2 \tag{25}$$

In this expression the field operator for the vector potential may be replaced by the matrix element

$$\langle \{\mathbf{k}, \varepsilon\} | e^{i\hat{H}_0 t'} \hat{\mathbf{A}} e^{-\hat{H}_0 t'} | 0 \rangle = \left( \frac{2\pi \hbar c}{k} \right)^{1/2} \frac{e^{i\hbar \omega t'} \varepsilon_{\mathbf{k}\varepsilon}}{L^{3/2}} \exp(-i\mathbf{k} \cdot \mathbf{x}) \tag{26}$$

With this replacement $H_{rad}$ becomes in effect an operator on the boson space only. The total energy radiated in time t is then on the average

$$E_{rad} = \sum_{\mathbf{k}, \varepsilon} P(k, \varepsilon) \hbar \omega, \quad \omega = kc \tag{27}$$

Substituting we find

$$E_{rad} = \sum_{k,\varepsilon} \hbar\omega \sum_{\Psi_{out}} \left| \langle \Psi_{out}, \{\mathbf{k},\varepsilon\} | \frac{1}{\hbar} \int_0^t dt' e^{i\hat{H}_0 t'} \hat{H}_{rad} e^{-i\hat{H}_0 t'} | \Psi_c, z, 0 \rangle \right|^2$$

$$= \sum_{k,\varepsilon} \hbar\omega \sum_{\Psi_{out}} \left| \langle \Psi_{out} | \frac{1}{\hbar} \int_0^t dt' e^{i\hat{H}_0 t'} \frac{-q}{mc} \int d^3x \hat{\Phi}^\dagger \left( \left(\frac{2\pi\hbar c}{k}\right)^{1/2} \frac{e^{i\hbar\omega t'} \varepsilon_{\mathbf{k}\varepsilon}}{L^{3/2}} \exp(-i\mathbf{k}\cdot\mathbf{x}) \right) \cdot (\mathbf{p} - \frac{q}{c}\mathbf{A}_E(x))\hat{\Phi} e^{-i\hat{H}_0 t'} | \Psi_c, z, 0 \rangle \right|^2 \quad (28)$$

Now we take the standard soft photon limit for weak external fields, where the exponential behavior is approximated as follows [2]:

$$\exp(-i\mathbf{k}\cdot\mathbf{x}) \approx 1 \quad (29)$$

then we obtain

$$E_{rad} = \sum_{k,\varepsilon} \hbar\omega \sum_{\Psi_{out}} \left| \langle \Psi_{out} | \frac{1}{\hbar} \int_0^t dt' e^{i\hat{H}_0 t'} \frac{-q}{mc} \int d^3x \hat{\Phi}^\dagger \left( \left(\frac{2\pi\hbar c}{k}\right)^{1/2} \frac{e^{i\hbar\omega t'} \varepsilon_{\mathbf{k}\varepsilon}}{L^{3/2}} \right) \cdot (\mathbf{p} - \frac{q}{c}\mathbf{A}_E)\hat{\Phi} e^{-i\hat{H}_0 t'} | \Psi_c, z, 0 \rangle \right|^2$$

$$= \sum_{k,\varepsilon} \hbar\omega \sum_{\Psi_{out}} \left| \langle \Psi_{out} | \frac{1}{\hbar} \int_0^t dt' e^{i\hat{H}_0 t'} \frac{-q}{mc} \left( \left(\frac{2\pi\hbar c}{k}\right)^{1/2} \frac{e^{i\hbar\omega t'} \varepsilon_{\mathbf{k}\varepsilon}}{L^{3/2}} \right) \cdot \left( \int d^3x \hat{\Phi}^\dagger (\mathbf{p} - \frac{q}{c}\mathbf{A}_E)\hat{\Phi} \right) e^{-i\hat{H}_0 t'} | \Psi_c, z, 0 \rangle \right|^2 \quad (30)$$

Defining the velocity operator by

$$\hat{\mathbf{V}} = \int d^3x \hat{\Phi}^\dagger (\mathbf{p} - \frac{q}{c}\mathbf{A}_E)/m \hat{\Phi} \quad (31)$$

then

$$E_{rad} = \sum_{\mathbf{k},\varepsilon} \hbar\omega \sum_{\Psi_{out}} \left| \langle \Psi_{out} | \frac{1}{\hbar} \int_0^t dt' e^{i\hat{H}_0 t'} \frac{-q}{c} \left( \left(\frac{2\pi\hbar c}{k}\right)^{1/2} \frac{e^{i\hbar\omega t'} \varepsilon_{\mathbf{k}\varepsilon}}{L^{3/2}} \right) \cdot \hat{\mathbf{V}} e^{-i\hat{H}_0 t'} | \Psi_c, z, 0 \rangle \right|^2 \quad (32)$$

Now we use the identity for orthogonal polarization basis vectors

$$\sum_\varepsilon \varepsilon^*_{\mathbf{k}\varepsilon,i}\varepsilon_{\mathbf{k}\varepsilon,j} = \delta_{i,j} - \frac{k_i k_j}{k^2} \qquad (33)$$

To obtain

$$E_{rad} = \sum_{\mathbf{k},i,j} \left(\frac{q^2 2\pi}{L^3}\right)\left[\delta_{i,j} - \frac{k_i k_j}{k^2}\right] \times$$
$$\left(\langle \Psi_c, z, 0 | \left(\int_0^t dt' e^{i\hat{H}_0 t'} e^{-i\hbar\omega t'} \hat{\mathbf{V}}_i e^{-i\hat{H}_0 t'}\right) \cdot \left(\int_0^t dt'' e^{i\hat{H}_0 t''} e^{i\hbar\omega t''} \hat{\mathbf{V}}_j e^{-i\hat{H}_0 t''}\right) | \Psi_c, z, 0\rangle\right) \qquad (34)$$

Now use

$$\sum_\mathbf{k} \to \int \frac{L^3 \omega^2}{8\pi^3 c^3} \sin(\theta) d\theta d\phi d\omega \qquad (35)$$

to get

$$E_{rad} = \int \frac{L^3 \omega^2}{8\pi^3 c^3}\sin(\theta)d\theta d\phi d\omega \left(\frac{q^2 2\pi}{L^3}\right)\sum_{i,j}\left[\delta_{i,j} - \frac{k_i k_j}{k^2}\right]\times$$
$$\left(\langle \Psi_c, z, 0 | \int_0^t dt' \int_0^t dt'' e^{iH_0 t'} e^{-i\hbar\omega(t'-t'')}\hat{\mathbf{V}}_i e^{-iH_0 t'} e^{iH_0 t''} \hat{\mathbf{V}}_j e^{-iH_0 t''} | \Psi_c, z, 0 \rangle\right) \qquad (36)$$

And now use

$$\int d\omega \omega^2 e^{iw(t-t')} = \pi \frac{\partial}{\partial t}\frac{\partial}{\partial t'}\delta(t-t') \qquad (37)$$

To obtain

$$E_{rad} = \int \frac{L^3}{8\pi^3 c^3} \sin(\theta) d\theta d\phi \left( \frac{q^2 2\pi}{L^3} \right) \sum_{i,j} \left[ \delta_{i,j} - \frac{k_i k_j}{k^2} \right] \times$$
$$\left( \left\langle \Psi_c, z, 0 \left| \int_0^t dt' \int_0^t dt'' e^{iH_0 t'} \pi \frac{\partial}{\partial t''} \frac{\partial}{\partial t'} \delta(t''-t') \hat{V}_i e^{-iH_0 t'} e^{iH_0 t''} \hat{V}_j e^{-iH_0 t''} \right| \Psi_c, z, 0 \right\rangle \right)$$
(38)

Define the time dependent velocity operator by

$$\hat{V}(t) = e^{iH_0 t'} \int d^3 x \hat{\Phi}^\dagger (\mathbf{p} - \frac{q}{c} \mathbf{A}_E) / m \hat{\Phi} e^{-iH_0 t'}$$
(39)

And substitute to obtain

$$E_{rad} = \int \frac{L^3}{8\pi^3 c^3} \sin(\theta) d\theta d\phi \left( \frac{q^2 2\pi}{L^3} \right) \sum_{i,j} \left[ \delta_{i,j} - \frac{k_i k_j}{k^2} \right] \int_0^t dt' \int_0^t dt'' \pi \frac{\partial}{\partial t''} \frac{\partial}{\partial t'} \delta(t-t') \times$$
$$\left\langle \Psi_c, z, 0 \left| \hat{V}_i(t') \hat{V}_j(t'') \right| \Psi_c, z, 0 \right\rangle$$
(40)

Integrating by parts twice yields

$$E_{rad} = \int \frac{L^3}{8\pi^2 c^3} \sin(\theta) d\theta d\phi \left( \frac{q^2 2\pi}{L^3} \right) \sum_{i,j} \left[ \delta_{i,j} - \frac{k_i k_j}{k^2} \right] \int_0^t dt'$$
$$\left\langle \Psi_c, z, 0 \left| \dot{\hat{V}}_i(t') \dot{\hat{V}}_j(t') \right| \Psi_c, z, 0 \right\rangle$$
(41)

This is starting to look like Larmor's formula. Performing the i, j sum we find

$$E_{rad} = -\int \frac{q^2}{4\pi c^3} \sin(\theta) d\theta d\phi \int_0^t dt'$$
$$\left\langle \Psi_c, z, 0 \left| \dot{\hat{V}}(t')^2 - \left( \dot{\hat{V}}(t') \cdot \hat{r} \right)^2 \right| \Psi_c, z, 0 \right\rangle$$
(42)

But this integrates out to

$$E_{rad} = \frac{2}{3} \frac{q^2}{c^3} \int_0^t dt' \left\langle \Psi_c, z, 0 \left| \dot{\hat{V}}(t')^2 \right| \Psi_c, z, 0 \right\rangle$$
(43)

Which is very similar to the classical Larmor's formula. We must still evaluate this for the coherent state though. Let's transform back to the Schrodinger representation:

$$E_{rad} = \frac{2}{3}\frac{q^2}{c^3}\int_0^t dt' \langle \Psi_c, z, t | \dot{\hat{\mathbf{V}}}^2 | \Psi_c, z, t \rangle \tag{44}$$

defining

$$\dot{\mathbf{v}} = \frac{i}{\hbar}\left[H_0, (\mathbf{p} - \frac{q}{c}\mathbf{A}_E)\right] \tag{45}$$

Then it is straightforward to show that

$$\dot{\hat{\mathbf{V}}} = \int d^3x \hat{\Phi}^\dagger(x)\dot{\mathbf{v}}\hat{\Phi}(x) = \int d^3x \hat{\Phi}^\dagger(x)\left[H_0, \frac{i}{\hbar}(\mathbf{p}-\frac{q}{c}\mathbf{A}_E)\right]\hat{\Phi}(x) \tag{46}$$

And therefore

$$E_{rad} = \frac{2}{3}\frac{q^2}{c^3}\int_0^t dt' \langle \Psi_c, z, t | \left(\int d^3x \hat{\Phi}^\dagger(x)\dot{\mathbf{v}}(x)\hat{\Phi}(x)\right)^2 | \Psi_c, z, t \rangle \tag{47}$$

Rewriting we get

$$E_{rad} = \frac{2}{3}\frac{q^2}{c^3}\int_0^t dt' \int d^3x \int d^3x' \times \\ \langle \Psi_c, z, t | \hat{\Phi}^\dagger(x)\dot{\mathbf{v}}(x)\hat{\Phi}(x)\hat{\Phi}^\dagger(x') \cdot \dot{\mathbf{v}}(x')\hat{\Phi}(x') | \Psi_c, z, t \rangle \tag{48}$$

But we have the relation

$$\hat{\Phi}^\dagger(x)\hat{\Phi}(x)\hat{\Phi}^\dagger(x')\hat{\Phi}(x') = \hat{\Phi}^\dagger(x)\hat{\Phi}^\dagger(x')\hat{\Phi}(x)\hat{\Phi}(x') + \hat{\Phi}^\dagger(x)\left[\hat{\Phi}(x), \hat{\Phi}^\dagger(x')\right]\hat{\Phi}(x') \\ = \hat{\Phi}^\dagger(x)\hat{\Phi}^\dagger(x')\hat{\Phi}(x)\hat{\Phi}(x') + \hat{\Phi}^\dagger(x)\hat{\Phi}(x')\delta^3(x-x') \tag{49}$$

And so

$$E_{rad} = \frac{2}{3}\frac{q^2}{c^3}\int_0^t dt' \int d^3x d^3x' \langle \Psi_c, z, t | \hat{\Phi}^\dagger(x)\hat{\Phi}^\dagger(x')\dot{\mathbf{v}}(x) \cdot \dot{\mathbf{v}}(x')\hat{\Phi}(x)\hat{\Phi}(x') | \Psi_c, z, t \rangle + \\ + \frac{2}{3}\frac{q^2}{c^3}\int_0^t dt' \int d^3x \langle \Psi_c, z, t | \hat{\Phi}^\dagger(x)\dot{\mathbf{v}}(x)^2\hat{\Phi}(x) | \Psi_c, z, t \rangle \tag{50}$$

Now we use the fact that the coherent state is an eigenstate for the field (11) to obtain the final result

$$E_{rad} = \frac{2}{3}\frac{q^2 \bar{n}^2}{c^3}\int_0^t dt' \left[\int d^3x \Psi^*(x,t')\dot{\mathbf{v}}(x)\Psi(x,t')\right]^2$$
$$+ \frac{2}{3}\frac{q^2 \bar{n}}{c^3}\int_0^t dt' \int d^3x \Psi^*(x,t')\dot{\mathbf{v}}(x)^2 \Psi(x,t') \qquad (51)$$

This can be calculated with the aid of the Lorentz force equation

$$m\dot{\mathbf{v}} = \frac{q}{2c}[\mathbf{v}\times\mathbf{B} - \mathbf{B}\times\mathbf{v}] - \nabla V(x) - c\nabla|\Psi(x)|^2 \qquad (52)$$

## 3. EXPERIMENTAL IMPLICATIONS

The two terms in (51) have a simple interpretation. The first is a hydrodynamic term. It is the formula one would get from a classical charged fluid [12]. It grows as the number of particles squared and so for large numbers it dominates in general. It involves the expectation value of the acceleration $|\langle\Psi|\mathbf{a}|\Psi\rangle|^2$ quantity squared. The second term in (51) is what one would have if the bosons were well separated. It depends on the expectation value of the acceleration squared $\langle\Psi|\mathbf{a}^2|\Psi\rangle$.

If the particle has an extended wave packet, then the results can differ substantially from the classical result for a point particle. For example, consider the situation in Figure 1, where the mean Schrödinger wave packet is moving in the x direction and is elongated in the x direction, ie. the x momentum uncertainty is very small. Let the force field which this particle is moving through be localized in a volume whose x dimension δ is much smaller than the wave packet length L.

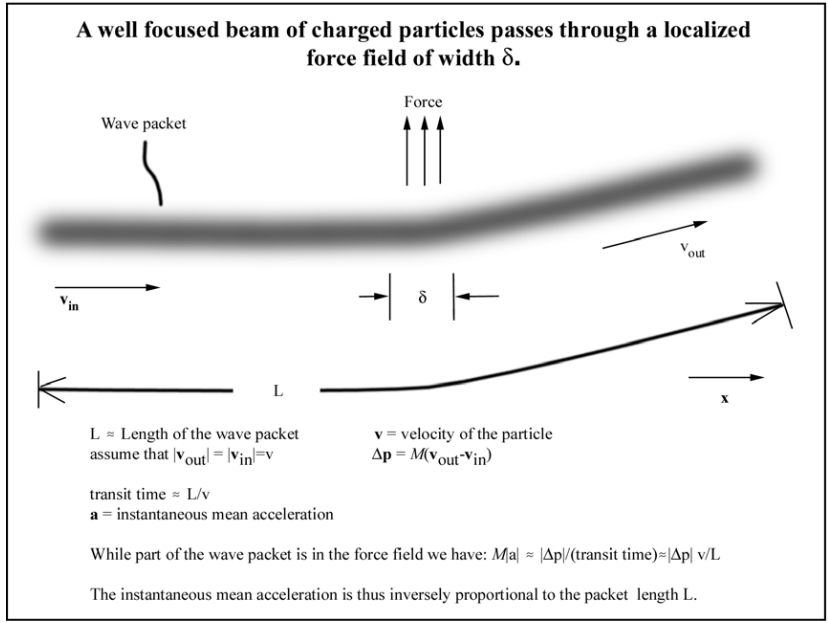

Figure 1 Localized force acting on an extended wave packet

Imagine that the wave packet in figure 1 is essentially invariant in shape along its length. We see that the instantaneous mean acceleration gets smaller inversely proportional to the wave packet's length while the impulse to the particle remains constant. The force is active on the wave packet for a time equal to the transit time, and so the total energy is the power integrated over this time.

Consider the first term in (51). The total energy radiated will then be, for values of L which are large compared with δ

$$\text{Total Energy radiated} \approx \frac{2}{3}\frac{q^2 \bar{n}^2}{c^3}|\langle\Psi|\mathbf{a}|\Psi\rangle|^2 (L/v)$$
$$\approx \frac{2}{3}\frac{q^2 \bar{n}^2}{c^3}\left(\frac{\Delta p \cdot \mathbf{v}}{m \cdot L}\right)^2 (L/v) \tag{53}$$

This expression goes to zero as 1/L in the limit of large L. Even though the total impulse imparted to the particle by the force is held constant, the total radiated energy goes to zero for very long wave packets. This is the effect that is being predicted here for the dominant term. It says that by preparing very long wave packets, bremsstrahlung can be suppressed and condensates can be accelerated with greatly reduced radiation.

The second term does not depend on L and so it represents radiation that cannot be suppressed, at least to lowest order perturbation theory.

What are the prospects of further suppression of radiation? It is clear that the effective coupling constant in the radiation calculation is growing with the number of particles, and so at some point the lowest order approximation will not be sufficient. Electromagnetic screening effects have already been studied for Bose-Einstein condensates [31]. If screening occurs, we can expect that radiation will have a high probability of being reabsorbed by the condensate. Therefore it can be expected that higher order corrections will further suppress the radiation, perhaps even leading to a kind of superconductivity. This could allow low loss charged particle storage beams which would have practical applications. Information could be stored in the condensate in the form of its mean wave function, and even energy might be stored. Observing the radiation from charged condensate beams could be a means of determining the shape and structure of the wave function.

**4.0 CONCLUSION**

Bremsstrahlung from Bose-Einstein charged condensates shows more dependence on the form of the wave function that does an incoherent beam of charged particles. The results are shown in Table 1.

| Classical Radiation Result | $E_{rad} = \frac{2}{3}\frac{q^2}{c^3}\int_0^T \mathbf{a}^2(t')dt'$ |
|---|---|
| Hydrodynamic Model Result | $E_{rad} = \frac{2}{3}\frac{q^2}{c^3}\int_0^T \left|\langle\Psi|\mathbf{a}(t')|\Psi\rangle\right|^2 dt'$ |
| Conventional Quantum Radiation Result for a single particle wave function | $E_{rad} = \frac{2q^2}{3c^3}\int_0^T \langle\Psi|\mathbf{a}^2(t')|\Psi\rangle dt'$ |
| Radiation from a Bose-Einstein condensate. Particles have charge q, and the mean number of particles is $\bar{n}$ | $E_{rad} = \frac{2}{3}\frac{q^2\bar{n}^2}{c^3}\int_0^T \left|\langle\Psi|\mathbf{a}(t')|\Psi\rangle\right|^2 dt'$ $+ \frac{2q^2\bar{n}}{3c^3}\int_0^T \langle\Psi|\mathbf{a}^2(t')|\Psi\rangle dt'$ |

Table I. Summary of results

The results show that the condensate acts mostly like a hydrodynamic model type of quantum fluid in the sense of [12], but that there is still a term that acts like a set of isolated bosons. When the acceleration occurs over only a small region the radiation can be greatly suppressed. Inclusion of higher order corrections might suppress this radiation further. In particular, reabsorption of already emitted photons by the condensate might become an important mechanism for long coherence length and large wave packets.

**ACKNOWLEDGEMENTS**

The author acknowledges valuable discussions with Vladimir Kresin during the course of this work.